\documentclass[aps,prd,noshowpacs,superscriptaddress,amsmath,amssymb,twocolumn,preprintnumbers,reprint]{revtex4}
\usepackage{amsmath,amssymb,lmodern}
\usepackage{bbold}
\usepackage{bm}
\usepackage{relsize}
\usepackage{braket}
\usepackage{bigints}

\usepackage{graphicx}
\graphicspath{{figures/}}

\renewcommand{\Re}{\operatorname{Re}}

\renewcommand{\v}[0]{\bm}

\renewcommand{\tt}{\textit}

\begin{document}
\title{Intensity of gluon bremsstrahlung in a finite plasma}
\author{X. Feal}
\email{xabier.feal@igfae.usc.es}
\author{R.A. Vazquez}
\email{vazquez@igfae.usc.es}
\affiliation{ 
Instituto Galego de F\'{\i}sica de Altas Enerx\'{\i}as \& \\
Departamento de F\'{\i}sica de Part\'{\i}culas \\
Universidade de Santiago de Compostela, 15782 Santiago, SPAIN
}

\date{\today}
\begin{abstract}
The intensity of single gluon bremsstrahlung in a QCD plasma is evaluated with
a Monte Carlo which solves the transport equation for a generic interaction
with the medium. In particular the calculation is performed for a Debye
screened potential and compared to the well known Gaussian/Fokker-Planck
results. The full calculation including the first and last gluons show 
a qualitatively different behavior from the BDMPS result for
any finite medium length. It is shown that the emission intensity is
underestimated for the Gaussian approximation, compared to the Debye screened
potential. This change can not be accounted for by a redefinition of the
Gaussian parameter ($\hat q$).
\end{abstract}
\maketitle

\section{Introduction}

The formation of the quark-gluon plasma in high energy collisions is an open
problem in QCD and a systematic study of the dependence of the medium
characteristics on the energy loss of high energy partons and other probes of
the medium are needed to fully understand it.  The existence of multiple
sources of scattering in a medium severely affects the way in which quanta are
emitted from high energy particles. Local internal phases in the scattering
amplitudes, growing with the traveled distance, regulate the amount of
scatterers which can coherently participate into a single emission
element. This interference effect, known as Landau-Pomeranchuk-Migdal (LPM)
suppression \cite{termikaelian1953,landau1953a}, leads to a substantial
reduction of a radiation scenario naively described as an incoherent sum of
single Bethe-Heitler \cite{bethe1934} intensities.

Semi-infinite medium calculations of this phenomenon under the Fokker-Planck
approximation have been introduced for QED in \cite{landau1953b,migdal1956}
and subsequent extensions for QCD
\cite{rbaier1995,zakharov1996a,wiedemann2000a} have been developed.  
These results 
\cite{dokshitzer2001,rbaier2001,salgado2003,armesto2004,arleo2017} 
have been formulated to account 
for the energy-loss mechanism originating the depletion of high transverse
momentum particles in heavy ion collisions at RHIC and LHC due to a multiple
scattering process with a medium, an indicative sign of QGP formation
\cite{bjorken1982}. We have to note, however, that in the Fokker-Planck
approximation, which naturally emerges when the number of collisions is large,
the medium interaction is replaced by an effective one
\cite{rbaier1995,zakharov1996a}. This leads to a Gaussian momentum
distribution which is valid for not too large deviations from the typical
$p_t$ accumulated by the particle. For QCD plasmas, these conditions are not
necessarily fulfilled.  First, since media are never very large, the number of
collisions can not be large. Second, realistic massless or massive/screened
interactions \cite{gyulassy1994}, like the Coulomb or Debye potentials,
respectively, have long $p_t$ tails substantially enhancing the emission
intensity. The semi-infinite medium approximation, meanwhile, loses the
relevant dependence with the medium length $l$.  In this limit, the intensity
becomes proportional to $l$, which produces an infinite suppression in the
regime of vanishing phases and, therefore, the soft photon theorem
\cite{weinberg1965} is not observed. While this approximation is adequate for
large mediums it becomes critical for the energy-loss estimation at
proton-proton or low centrality heavy ion collisions.
Following these
concerns several frameworks beyond the Gaussian approximation have been
developed for finite QCD media
\cite{gyulassy1999,gyulassy2000,gyulassy2001}, nuclei
\cite{wang2001a,guo2006}, and for the case of heavy quarks
\cite{djordjevic2004,zhang2004}.
As expected these results have shown significant differences with the
semi-infinite length calculations and they have produced reliable predictions
at RHIC and LHC \cite{vitev2002,chang2014,djordjevic2015}.

Taking these conditions into account we develop a formalism which is able to
account for any interaction potential with the medium and incorporates gluon's
transverse spectrum dependence, if desired. In section \ref{radiation} we
briefly present the formalism for a general scenario, whereas in section
\ref{interior_radiation} we restrict the intensity evaluation to the spectrum
of gluons emitted after a first hard collision. Finally, we present the main
conclusions.

\section{Radiation intensity}
\label{radiation}
\begin{figure}[h]
\includegraphics[scale=0.41]{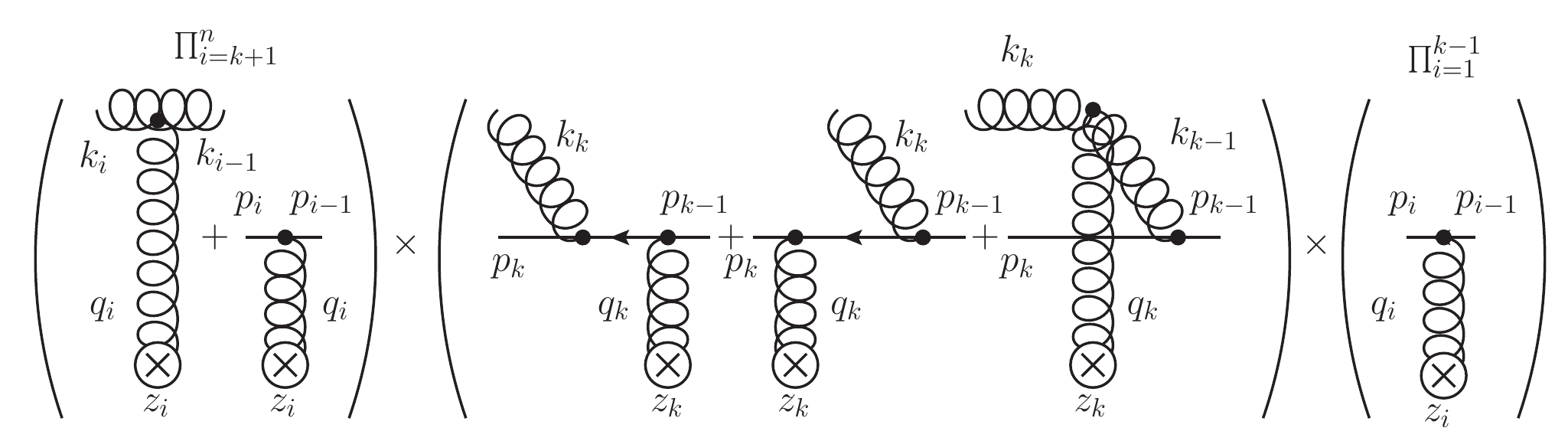}
\caption{Diagrammatic representation of the discretized
  Eq.\eqref{emission_amplitude_allterms}, at leading order in the coupling
  $g_s$, summation from $k=1$ to $k=n$ assumed.}
\label{fig:figure_0}
\end{figure}

We consider an ideal scenario in which an on-shell quark coming from
infinity undergoes a multiple scattering process in a finite medium, and then
goes to infinity. The amplitude of a single gluon emission can be written as
\begin{equation}
M = -i g_s \int d^4x  \bar{\Psi}_2(x)A_{\mu}^\dag(x) \gamma^\mu \Psi_1(x),
\label{emission_amplitude_general}
\end{equation}
where $\Psi_{1,2}(x)$ represent the quark wave function before and after the
emission, $A^\mu(x)= A^\mu_\alpha(x) t_\alpha$ represents the emitted colored
gluon, $t_{\alpha}$ are generators of SU(3) and $g_s=\sqrt{\alpha_s}$ is
the coupling constant.  Both quark and gluon are subject to the external field
of the medium, then at $x$ they are a superposition of a free state plus a set
of scattered states, of the form
\begin{equation}
\Psi_1(x) = \int  \frac{d^3\v{p}}{(2\pi)^3} S_q(p,p_0;z,0)\Psi(p,x),
\end{equation}
where $\Psi(p,x)= \mathcal{N}(p)u(p)e^{-ip\cdot x}$ is a free quark and
$\mathcal{N}(p)=\sqrt{m/p^0}$. Similarly for the gluon
\begin{equation}
A_\mu(x) = \int  \frac{d^3\v{k}}{(2\pi)^3} S_g(k,k_f;z,l) A_\mu^{(0)}(k,x).
\end{equation}
where $A_\mu^{(0)}(k,x)=\mathcal{N}(k) \epsilon_\mu(k) e^{-i k\cdot x}$ is a
free gluon, $\mathcal{N}(k)=\sqrt{2\pi/\omega}$ the normalization and
$\epsilon_\mu(k)$ its polarization. The amplitudes
$S_q(p_2,p_1;z_2,z_1)$ ($S_g(k_2,k_1;z_2,z_1)$) are beyond eikonal
evaluations of the elastic amplitudes for a quark (gluon) changing its momentum
$p_1\to p_2$ ($k_1\to k_2$), spin $s_1\to s_2$ (polarization $\lambda_1\to
\lambda_2$) and color $a_1\to a_2$ ($\alpha_1\to \alpha_2$) due to
the amount of matter between $z_1$ and $z_2$. They represent elastic interactions with the
medium, characterized by $n$ thin sheets of $n(z_i)$ scattering sources of
thickness $\delta z$ spanning a length $l$, of radius $R$, with density
denoted by $n_0(z_i)$, of the form \cite{gyulassy1994}
\begin{equation}
A^0_{\alpha_i}(x_i) = g_s t_{\alpha_i}\sum_{j=1}^{n(z_i)}\frac{1}{|\v{x}_i-\v{r}_j|}
e^{-\mu_d|\v{x}_i-\v{r}_j|},
\end{equation}
with a Debye mass $\mu_d$. Mixed scenarios adding gluons of the QGP can be
taken into account by redefining the coupling. In the high energy limit these
amplitudes are given by convolutions of the form
\begin{align}
S_q(p_n,p_0;z_n,z_1)=2\pi\delta(p_n^0-p_0^0)\delta^{s_n}_{s_0}\beta_p\left(\prod_{i=1}^{n-1}\int
\frac{d^2\v{p}_i^t}{(2\pi)^2}\right)\nonumber\\
\times\left(\prod_{i=1}^n\int d^2\v{x}_i^t
e^{-i\v{q}_i^q\cdot\v{x}_i}\exp\left[-i\frac{g_s}{\beta_p}t_{\alpha_i}\int
  dz A_{\alpha_i}^0(x_i)\right]\right)\label{elastic_amplitude_quark}
\end{align}
where $\v{q}_i^q\equiv \v{p}_i-\v{p}_{i-1}$ is the 3-momentum transfer of the
quark at the layer ($i$), $\beta_p$ its
velocity, and
\begin{align}
S_g(k_n,k_0;z_n,z_1)=2\pi\delta(\omega_n^0-\omega_0^0)\delta^{\lambda_n}_{\lambda_0}\beta_k\left(\prod_{i=1}^{n-1}\int
\frac{d^2\v{k}_i^t}{(2\pi)^2}\right)\nonumber\\
\times\left(\prod_{i=1}^n\int d^2\v{x}_i^t
e^{-i\v{q}_i^g\cdot\v{x}_i}\exp\left[-i\frac{g_s}{\beta_k}T_{\alpha_i}\int
  dz A_{\alpha_i}^0(x_i)\right]\right)\label{elastic_amplitude_gluon}
\end{align}
where $\v{q}_i^g\equiv \v{k}_i-\v{k}_{i-1}$ is the 3-momentum transfer of the
gluon at the layer ($i$), $\beta_k$ its
velocity and $T_{\alpha}$ the generators of the adjoint representation of
SU(3). Color indices are implicit so the above amplitudes do have matrix
structure and carry ordered longitudinal phases regulating the LPM effect. If the medium transverse
dimension verifies $R\gg r_d$, being $r_d = 1/\mu_d$ the dimensions of a
scattering center (the Debye radius of the plasma), the momentum change in a
coherent scattering after traversing a small length in the medium will verify
$q_{coh}\sim 1/R \ll \mu_d \sim q_{incoh}$. In that case, the emission is
dominated by incoherent averages. We can write for the emission amplitude
\begin{align}
 M(l)=g_s \mathcal{N}(\omega)\int_0^l dz &\int_{\v{p}(0)}^{\v{p}(l)}
\frac{\mathcal{D}^3\v{p}}{(2\pi)^3}\int_{\v{k}(z)}^{\v{k}(l)}
\frac{\mathcal{D}^3\v{k}}{(2\pi)^3} \label{emission_amplitude_allterms}\\
&\exp\bigg(+\frac{i}{E_q}\int^l_z  ds k_\mu(s) p^\mu(s)\bigg) \nonumber \\
\frac{d}{dz}\bigg\{\frac{\epsilon_\mu^\lambda
      (z)p^\mu(z)}{k_\mu(z) p^\mu(z)} &
\big(S_g^{el}(k(l),k(z);l,z)\big)_{\alpha_l\alpha} \nonumber \\
&\big(S_q^{el}(p(l),p(z);l,z)\big)_{a_l,a'}\big(t_\alpha\big)_{a'a} \nonumber \\
&\big(S_q^{el}(p(z)+k(z),p(0);z,0)\big)_{aa_0}\big)\bigg\}\nonumber,
\end{align}
where Greek and Latin indices run in gluon and quark color dimensions,
respectively, sum over repeated indices is assumed and $z$ is the point in
which the gluon is emitted. The elastic amplitudes $S_q^{el}$ and
$S_g^{el}$ in \eqref{emission_amplitude_allterms} are now eikonal and thus
given by doing $(q_i^q)_z=0$ and $(q_i^g)_z=0$ in \eqref{elastic_amplitude_quark} and
\eqref{elastic_amplitude_gluon}, respectively. A diagrammatic representation of this expression is
shown in Figure \ref{fig:figure_0}. Amplitude
(\ref{emission_amplitude_allterms}) agrees with the QED classical formula in momentum
space \cite{landau1953b} except for the color modifications and the gluon
rescattering.
In Eq.(\ref{emission_amplitude_allterms}) the medium discretization produced discretized
parton paths $\v{p}_i$ for the quark and $\v{k}_i$ for the gluon, for
$i=0,1,..,n$ steps. Here $\mathcal{D}^3\v{p}/(2 \pi)^3$ is a shorthand
notation for $d^3 \v{p}_1/(2\pi)^3\cdots d^3 \v{p}_{n-1}/(2 \pi)^3$. The square amplitude can
be averaged over the transverse medium coordinates of each sheet and
integrated in the final gluon solid angle, giving the
intensity of emission of a gluon within an energy interval from $\omega$ to
$\omega+d\omega$, per unit of medium transverse size, in a total length $l$
\begin{align}
\omega& \frac{dI(l)}{d\omega} = \alpha_s \frac{N^2_c-1}{2N_c} \beta^2_k\omega^2\int \frac{d\Omega_{k_n}}{(2\pi)^2}
\left(\prod_{k=0}^{n-1}\int 
\frac{d^3\v{k}_j}{(2\pi)^3} \right)\label{squared_emission_amplitude_allterms}\\ 
&\left(\prod_{k=1}^{n}\phi(\delta\v{k}_k,
\delta z) \right)\bigg(h^{n}(y) \left|\sum_i^{n} \v{\delta}_i^{n}
\right|^2+ h^{s}(y)\left|\sum_{i=1}^n  \delta_i^{s}\right|^2\bigg)\nonumber
\end{align}
where $y=\omega/E_q$ and the functions $h^{n}(y)$ and $h^{s}(y)$ are the
kinematical weights appearing in the diagonal and non-diagonal sum in spins
and helicities of the squared emission vertices, at $z_i$ and $z_j$.
In the high energy limit 
\begin{equation}
h^{n}(y) = \frac{1+(1-y)^2}{2}, \medspace\medspace\medspace h^{s}(y)=\frac{y^2}{2}.
\nonumber
\end{equation}
The $\v{\delta}_i^{n}$ can be interpreted as a classical current, 
representing spin preserving amplitudes and is given by
\begin{equation}
\v{\delta}_i^{n}= \v{p}(0) \times \left(\frac{\hat{\v{k}}(z_{i+1})}{k_\mu(z_{i+1})
  p^\mu(0)}-\frac{\hat{\v{k}}(z_{i})}{k_\mu(z_{i})p^\mu(0)}\right)
e^{i \varphi_i},
\label{classical_current}
\end{equation}
whereas the spin flipping amplitudes produce
\begin{equation}
\delta_i^{s}=
\bigg(\frac{p^0(0)}{k_\mu(z_{i+1})p^\mu(0)}-\frac{p^0(0)}{k_\mu(z_{i})p^\mu(0)} \bigg) 
e^{i \varphi_i},
\label{quantum_current}
\end{equation}
and $\varphi_i = 1/E_q\sum_{k=i}^{n} \delta z  k_\mu(z_k)p^\mu(0)$. Once integrated along $\v{k}(z)$ with the elastic weights, they produce two
contributions of the same order. The spin flipping part, however, only becomes
relevant for gluon energies $\omega$ of the order of $E_q$ due to the form of
$h^{s}(y)$, in accordance with the classical behavior at the infrared
divergence. In Eq. (\ref{squared_emission_amplitude_allterms}) we are assuming $\omega
\ll E_q$ and therefore, we will neglect the contribution of $h^{s}(y)$ in what
follows. Within the same approximation one can assume that the quark momenta
is frozen $p_\mu(z) \sim p_\mu(0)$.

The phases and denominators appearing above are
\begin{eqnarray}
\frac{k_\mu(z)p^\mu(0)}{E_q} = 
\left(1-\beta(\omega)\right)\omega
+ \frac{ \delta \v{k}^2(z_k)}{2\omega \beta(\omega)}
\approx \frac{m_g^2+\v{k}_t^2(z_k)}
{2\omega}.
\label{phase_definition}
\end{eqnarray}
The gluon's velocity is taken as $\beta^2(\omega)=1-m_g^2/\omega^2$, where a
plasma mass $m_g$ has been introduced to take into account possible medium
effects in the gluon dispersion relation.  The gluon effective mass can be
considered of the order of $\mu_d$ \cite{blaizot2002}.  The color matrices
factor out of the elastic amplitudes and they can be averaged independently
over color configurations at the vertex in $z$.  Finally, the elastic weights
in (\ref{squared_emission_amplitude_allterms}) are calculated by an incoherent average
of the elastic amplitudes squared, the dominant term is given by the factor
$S_gS_q$, which gives in a layer of thickness $\delta z$
\begin{align}
\nonumber
\phi(\delta \v{k},\delta z) = 
e^{-n_0 \delta z \sigma_{gq}}
(2\pi)^3\delta^3(\delta \v{k})+ \nonumber \\
2\pi\delta(\delta k^0) \Sigma_2(\delta\v{k},\delta z).
\label{local_elastic_weight}
\end{align}
The first contribution represents the no collision probability within the
layer of thickness $\delta z$ and density $n_0(z)$ times the forward
distribution. The quantity $\lambda_{gq}^{-1}=\sigma_{gq}n_0(z)$ is the mean
free path of the gluon, $\sigma_{gq}$ its elastic cross section with a single
scattering center. The second contribution represents the collisional
distribution in case of collision.  It can be shown to satisfy a Moliere's
transport equation
whose solution reads
\begin{equation}
\Sigma_2(\v{q},\delta z) = \int d^2\v{x}
e^{-i\v{q}\cdot\v{x}} 
e^{-n_0 \delta z \sigma_{gq}}
\left( 
e^{n_0(z)\delta z \sigma_{gq}(|\v{x}|)}
-1 \right),
\label{moliere_solution}
\end{equation}
and where the Fourier transform of the squared single elastic amplitude reads
at first order in the coupling
\begin{align}
\sigma_{gq}(|\v{x}|)=\frac{4\pi\alpha_s^2}{\beta^2(\omega)\mu_d^2}T_f
\mu_d| \v{x}|K_1(\mu_d|\v{x}|).
\label{gluon_single_cross_section2}
\end{align}
where $T_f = 1/2$ is the first order Casimir for SU(3).  Using
(\ref{gluon_single_cross_section2}) the single elastic cross section required
at (\ref{moliere_solution}) is given at leading order by
\[\sigma_{gq}
=\frac{4\pi\alpha_s^2}{\beta^2(\omega)\mu_d^2}T_f.
\]
The convolution of (\ref{local_elastic_weight}) over a step $\delta l$
produces rules for momentum additivity. For a step verifying $\delta l <
\lambda_{gq}=1/n_0\sigma_{gq}$ and constant density an expansion in small
$\delta l \lambda_{gq}^{-1}$ is enough and the total scattering distribution
reduces to the incoherent superposition of the single distributions for the
matter in $\delta l$.  The squared momentum change in $\delta l$ is equal to a
momentum change in a single scattering,
\begin{equation}
\left<\delta\v{k}^2(\delta l) \right> = \mu_d^2
\bigg(2\log\bigg(\frac{2\omega}{\mu_d}\bigg)-1\bigg)=
\mu_d^2 \; \eta(\omega), 
\label{single_momentum_change}
\end{equation}
where $\eta(\omega)$ accounts for the long tail correction of the Debye
potential. For arbitrary distances $l$, using equation
(\ref{moliere_solution}) it can be shown that the squared momentum change is
additive in the traveled length. Indeed
\begin{equation}
\frac{\partial}{\partial l} \left<\delta \v{k}^2(l)\right> =
n_0\sigma_{gq}\left<\delta\v{k}^2(\delta l) \right> \equiv 2\hat{q}, 
\label{multiple_momentum_change}
\end{equation}
where we have defined the transport coefficient $\hat{q}$. Since the single
momentum change depends on the gluon's energy, $\hat{q}$ has to be fixed, in
principle, for each gluon's energy.

Evaluation of (\ref{squared_emission_amplitude_allterms}) can be accomplished by taking
the $\delta z \to 0$ limit, either by using a Boltzmann transport equation or
equivalently by integrating the kinetic phase with the elastic weights
(\ref{local_elastic_weight}), producing a path integral. In both cases, by
using the Fokker-Planck/Gaussian approximation for (\ref{moliere_solution})
Migdal's result \cite{migdal1956} for QCD matter is found
\cite{rbaier1995,zakharov1996a,wiedemann2000a}.

We have built a Monte Carlo program where (\ref{squared_emission_amplitude_allterms})
is evaluated as a sum over discretized paths. At each step the potential
distribution is sampled and the gluon's path is built. The quark is also
allowed to interact with the medium (so that its momentum does change). Phases
are calculated with the exact kinematical expression.  In this way one can
calculate the gluon emission distribution for any given potential, in
particular it has been calculated for the Debye potential and for the Gaussian
approximation.  A similar Monte Carlo was made for QED \cite{feal2018} whose
results reproduce the path integral limit for the Gaussian potential and agree
with Migdal's expression for $l\to\infty$, although our results are valid for
an arbitrary size of the medium. In a typical run, the step size is taken as
$0.01 \lambda_{gq}$, so that for medium size of $l \sim 5$ fm we have $\gtrsim
10^4$ steps. Paths are calculated for an array of $\sim 100 $ frequencies of
the gluon and $\sim 800 $ different emission angles. For these values, we run
$\approx 10^4$ simulations and average over them. This takes $\approx 24 $ h of
CPU time in a PC. We have checked that the size of the grid and the number of
simulations are enough for a statistical uncertainty less than $10 \%$ in all
cases. 
\begin{figure}[h]
\includegraphics[scale=0.65]{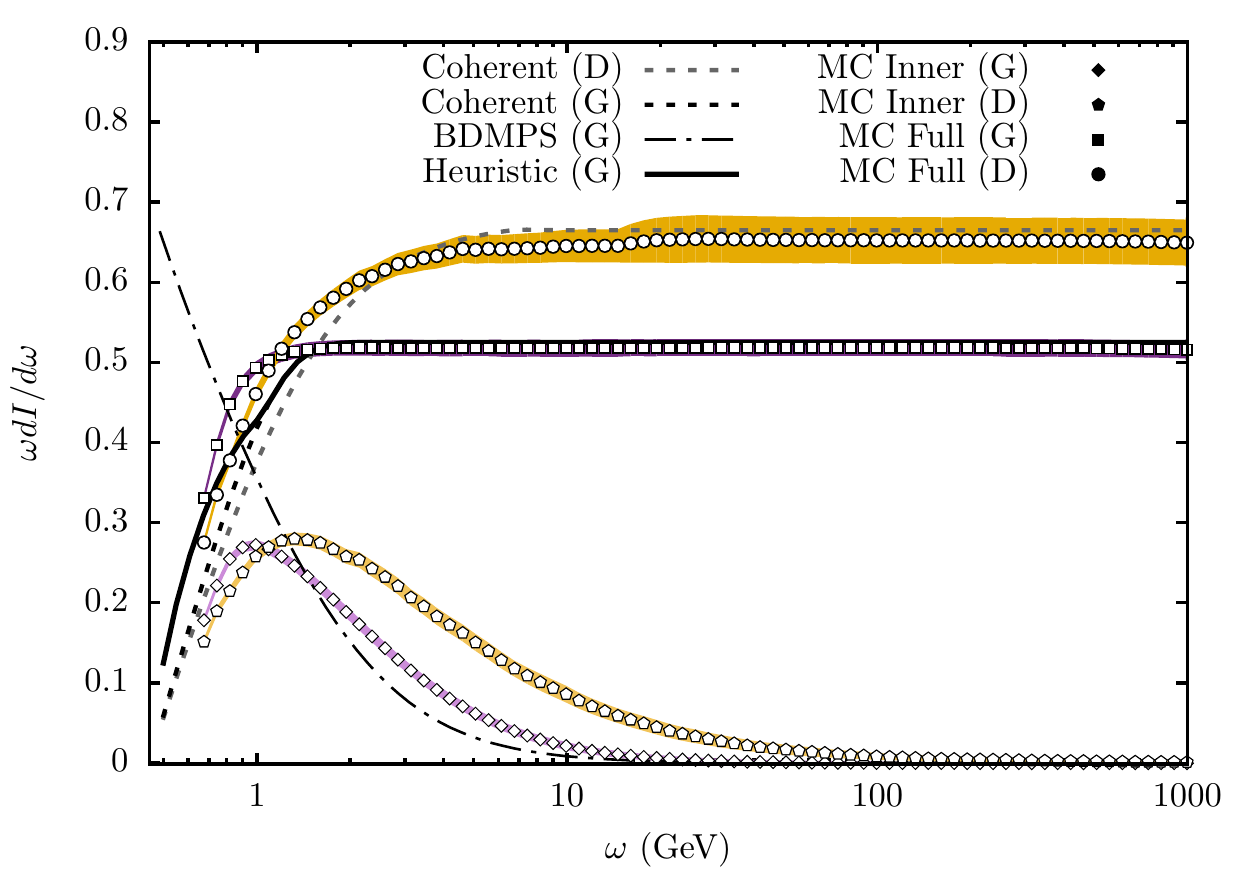}
\caption{Differential spectrum for gluons of $m_g$ = 0.45 GeV for a medium of
  $l=$ 1 fm and $n_0T_f$ = 8 fm$^{-3}$ (equivalently $\hat{q}$ = 1 GeV$^2/$fm). We show
 calculations for our Monte Carlo results both for the Debye potential (D), using
 the total intensity \eqref{squared_emission_amplitude_allterms} (circles) and
 the inner intensity \eqref{squared_emission_amplitude_innerterms} (pentagons), and
 for the Fokker-Planck (G) approximation, using the total intensity (squares) and the
 inner intensity (diamonds). Also shown is our approximation
 \eqref{simplificada} (solid line) and the 
 BDMPS result (dot-dashed line). The coherent limits both for the Fokker-Planck
 approximation (dotted dark line) and the Debye potential (dotted light line)
 are also shown.}
\label{fig:figure_1}
\end{figure}
\begin{figure}[h]
\includegraphics[scale=0.65]{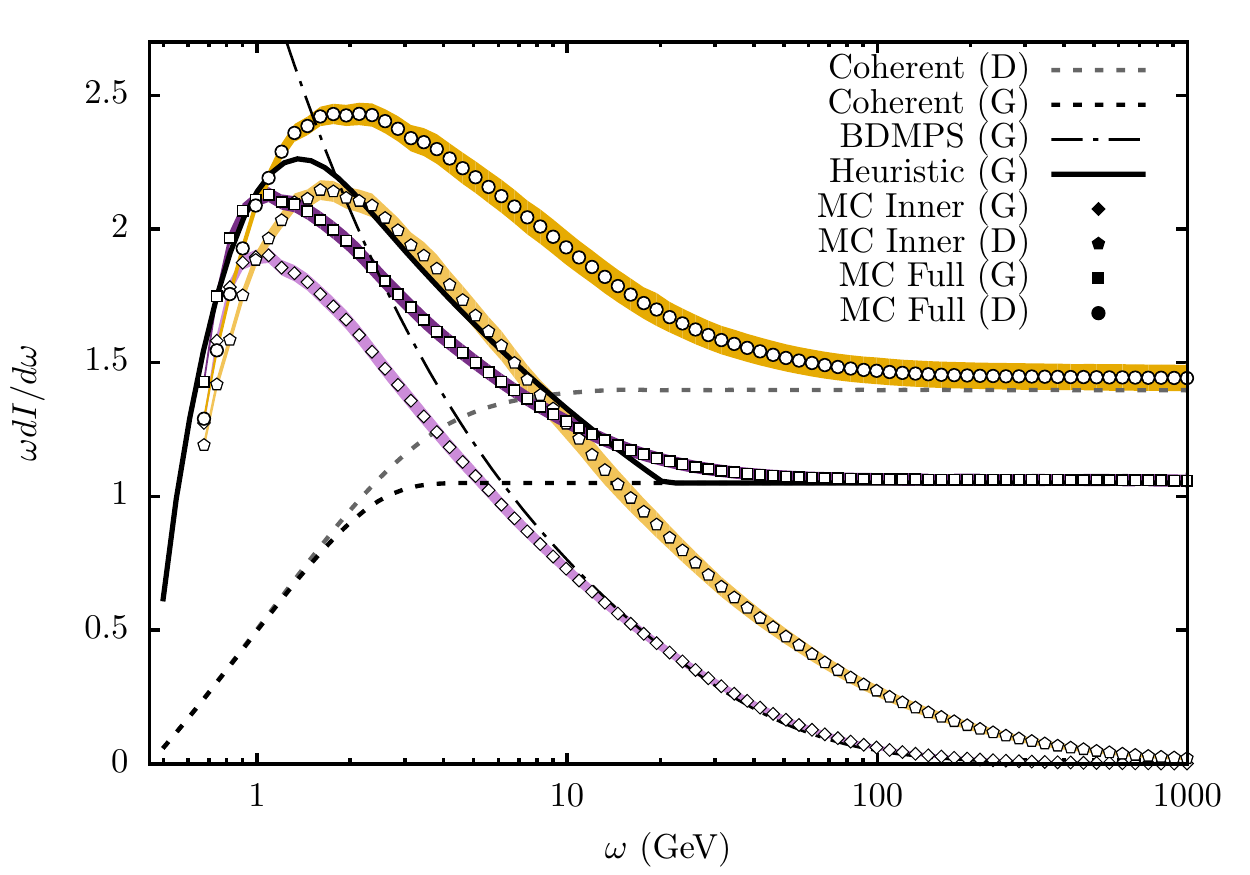}
\caption{Differential spectrum for gluons of $m_g$ = 0.45 GeV for a medium of
  $l$ = 5 fm and $n_0T_f$ = 8 fm$^{-3}$ (equivalently $\hat{q}$ = 1 GeV$^2/$fm). We show
 calculations for our Monte Carlo results both for the Debye potential (D), using
 the total intensity \eqref{squared_emission_amplitude_allterms} (circles) and
 the inner intensity \eqref{squared_emission_amplitude_innerterms} (pentagons), and
 for the Fokker-Planck (G) approximation, using the total intensity (squares) and the
 inner intensity (diamonds). Also shown is our approximation
 \eqref{simplificada} (solid line) and the 
 BDMPS result (dot-dashed line). The coherent limits both for the Fokker-Planck
 approximation (dotted dark line) and the Debye potential (dotted light line)
 are also shown.}
\label{fig:figure_2}
\end{figure}
On the other hand, we can take a simpler approach if we note that the single
radiation elements appearing in the sums (\ref{classical_current}) and
(\ref{quantum_current}) are interpretable as single Bethe-Heitler amplitudes
for a gluon emitted at $z_i$ due to the medium in $\delta z$, with a phase
which produces interferences in the squared amplitude. Each diagram appears
twice, representing the possibility that the gluon can be emitted either after
the $(i)$th change of momentum or just before the $(i+1)$th, except those two
cases where gluons are emitted before and after the first and last available
momentum changes. When the medium is removed, $l=0$, following
(\ref{local_elastic_weight}), momentum homogeneity $k(z)=k(l)$ cancels the
sums (\ref{classical_current}) and (\ref{quantum_current}) and the radiation
vanishes. For small media $l \lesssim \lambda_{gq}$ internal sum cancels and
we are left with the first and the last terms only
which reproduces the single Bethe-Heitler spectrum for a
$qq\to qqg$ process. By taking the limit $m_g\to 0$ and neglecting the spin
flip suppressed part, the Bertsch-Gunion formula is recovered
\cite{gunion1982}.  For arbitrary larger $l$, the internal structure in the
multiple scattering $k(z)$ becomes relevant and the sum in
(\ref{classical_current}) and (\ref{quantum_current}) is modulated by the
phase. We can pair terms in (\ref{classical_current}) and
(\ref{quantum_current}) in groups separated by a maximum distance $\delta
l=z_j-z_i$ having a relative phase (\ref{phase_definition}) under the elastic
weight (\ref{local_elastic_weight}), of the order of $1$, a condition which
reads
\begin{equation}
\varphi^{j}_i\equiv \frac{1}{E_q}\int^{z_j}_{z_i}dz k_\mu(z) p^\mu(0) \approx
\frac{m_g^2}{2w}\delta l +\frac{\hat{q}}{2\omega}(\delta l)^2=1.
\label{coherence_condition}
\end{equation} 
In each group, the partial internal sum between $z_i$ and $z_j$ at
(\ref{classical_current}) or (\ref{quantum_current}) cancels, since their
relative phase is negligible using condition (\ref{coherence_condition}). The
elements in the group act coherently between themselves, but they incoherently
interfere with any other group due to condition
(\ref{coherence_condition}). Using (\ref{coherence_condition}) this defines a
coherence length modulated by $\omega$, we call $\delta l(\omega)$, hence
\begin{equation}
\delta l(\omega) =
\frac{m_g^2}{2\hat{q}}\bigg(\sqrt{1+\frac{8\hat{q}\omega}{m_g^4}}-1\bigg),
\; \; \; \omega\le\omega_c,
\label{coherence_lenght}
\end{equation}
and $\delta l(\omega)=l$ for $\omega>\omega_c$, where $\omega_c$ verifies
$l(\omega_c)=l$, {\it i.e.} $\omega_c\sim(\hat{q}l+m_g^2)l$.  Over the length
$\delta l(\omega)$ the scattering centers are not resolved due to the negligible
accumulated phase change. This part of the medium acts like a single
scatterer of an equivalent charge $n_0\delta l(\omega)$.
Since there are $l/\delta l(\omega)$ of these partial sums acting
incoherently, we can write for the total intensity
\begin{align}
\nonumber
\omega\frac{dI(l)}{d\omega}=\frac{l}{\delta l(\omega)}\alpha_sC_f \beta^2(\omega)\omega^2\int 
\frac{d^2\Omega_{k_1}}{(2\pi)^2}\int\frac{d^2\Omega_{k_0}}{(2\pi)^2} \nonumber \\
\bigg(h^{n}(y)\big|\v{\delta}_1^{n}\big|^2+h^{s}(y)\big|\delta_1^{s}\big|^2
\bigg)\Sigma_2(\delta\v{k}(z_1),\delta l(\omega)),
\label{spectrum}
\end{align}
The above equation is a very good approximation in the frequency interval
$(m_g,\omega_s)$ in which the gluon completely resolves each of the single
scattering centers and in the interval $(\omega_c,E_q)$ in which the gluon
stops being able to resolve any internal structure of the medium. These values
are given by $\delta l(\omega_s)=\lambda_{gq}$ and $\delta l(\omega_c)=l$,
their phases satisfying $\varphi_0^{\lambda}(\omega_c) = 1$ and
$\varphi_0^{l}(\omega_s)=1$. Using (\ref{coherence_lenght})
$\omega_s=m_g^4/\hat{q}$.  For $\omega\gg \omega_c$ the entire medium acts
coherently like a single scatterer with an equivalent charge contained in the
length $l$ following a Bethe-Heitler power law $1/\omega$. Radiation intensity
in this interval depends then on the medium length and energy loss is
dominated by these gluons. For frequencies below $\omega_c$ gluon resolution
power starts to decouple the medium in groups of charges of big size, so
radiation grows as $l/\delta l(\omega)\sim\sqrt{\hat{q}/\omega}$ times a
Bethe-Heitler power law $1/\omega$, with a slow logarithmic charge decrease
$\omega/\hat{q}$. This decoupling saturates at $\omega_s$, when the coherence
length acquires the minimum length required to produce radiation, given by
$\lambda_{gq}$. However, suppression due to a vanishing velocity $\beta(\omega)$
rapidly cancels the enhancement. Using (\ref{spectrum}) we then find for the
intensity produced after traversing a length $ l$
\begin{equation}
\omega \frac{dI(l)}{d\omega} =\frac{\alpha_sC_f}{\pi^2}
\frac{l}{\delta l(\omega)} \int_0^\pi
d\theta \sin(\theta) F(\theta)\Sigma_2(\delta\v{k},\delta
l(\omega)),
\label{simplificada}
\end{equation}
with $|\delta\v{k}|=2\beta(\omega)\omega\sin(\theta/2)$ and
\begin{eqnarray}
\nonumber
F(\theta)=\bigg[\frac{1-\beta^2(\omega)\cos\theta}{2\beta(\omega)\sin(\theta/2)
\sqrt{1-\beta^2(\omega)\cos^2(\theta/2)}} \nonumber \\
\log\bigg[\frac{\sqrt{1-\beta^2(\omega)
\cos^2(\theta/2)}+\beta(\omega)\sin(\theta/2)}{\sqrt{1-\beta^2(\omega)
\cos^2(\theta/2)}-\beta(\omega)\sin(\theta/2)}\bigg]-1\bigg].
\end{eqnarray}
\begin{figure}[h]
\includegraphics[scale=0.65]{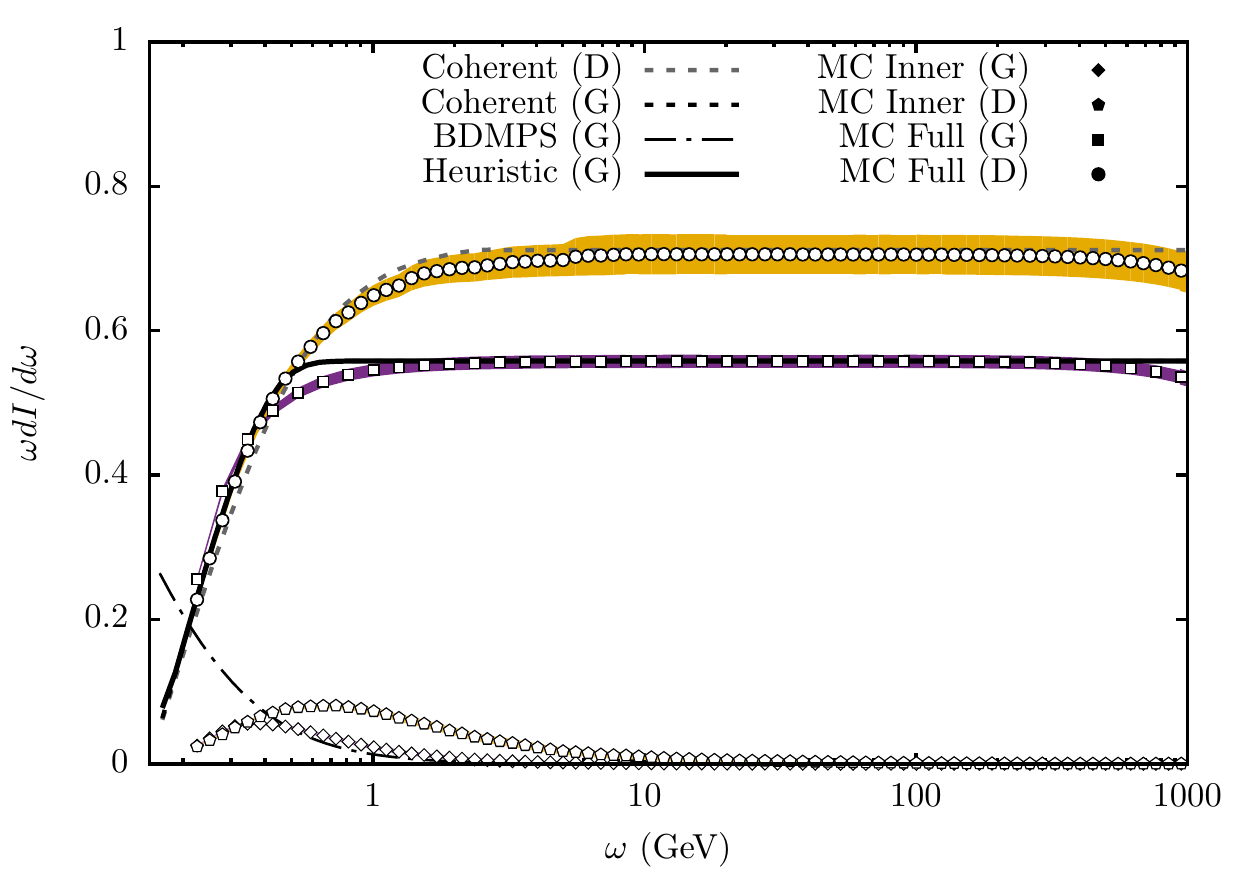}
\caption{Differential spectrum for gluons of $m_g$ = 0.15 GeV for a medium of
  $l$ = 1 fm and $n_0T_f$ = 1 fm$^{-3}$ (equivalently $\hat{q}$ = 0.12 GeV$^2/$fm). We show
 calculations for our Monte Carlo results both for the Debye (D) potential, using
 the total intensity \eqref{squared_emission_amplitude_allterms} (circles) and
 the inner intensity \eqref{squared_emission_amplitude_innerterms} (pentagons), and
 for the Fokker-Planck (G) approximation, using the total intensity (squares) and the
 inner intensity (diamonds). Also shown is our approximation
 \eqref{simplificada} (solid line) and the 
 BDMPS result (dot-dashed line). The coherent limits both for the Fokker-Planck
 approximation (dotted dark line) and the Debye potential (dotted light line)
 are also shown.}
\label{fig:figure_3}
\end{figure}
\begin{figure}[h]
\includegraphics[scale=0.65]{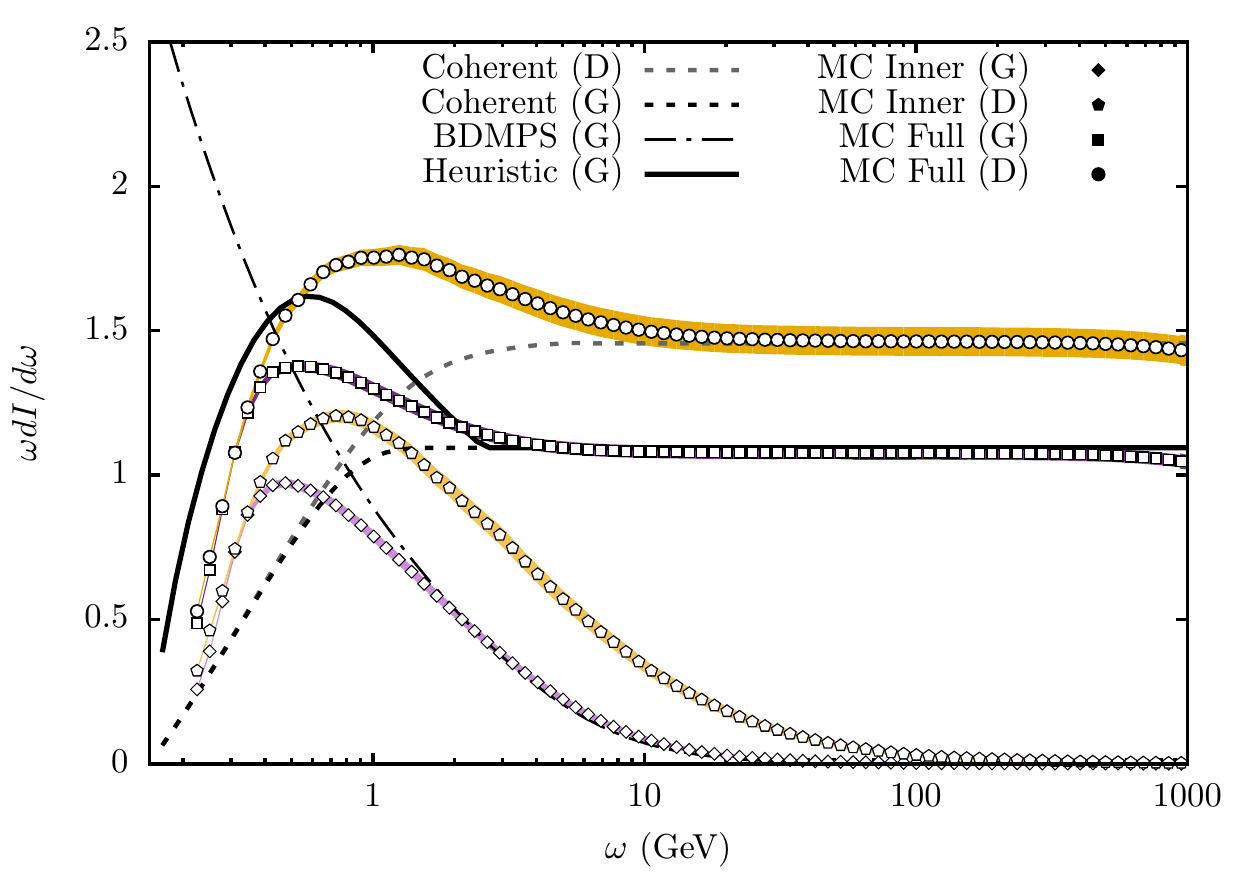}
\caption{Differential spectrum for gluons of $m_g$ = 0.15 GeV for a medium of
  $l$ = 5 fm 
 and $n_0T_f$ = 1 fm$^{-3}$ (equivalently $\hat{q}$ = 0.12 GeV$^2/$fm). We show
 calculations for our Monte Carlo results both for the Debye (D) potential, using
 the total intensity \eqref{squared_emission_amplitude_allterms} (circles) and
 the inner intensity \eqref{squared_emission_amplitude_innerterms} (pentagons), and
 for the Fokker-Planck (G) approximation, using the total intensity (squares) and the
 inner intensity (diamonds). Also shown is our approximation
 \eqref{simplificada} (solid line) and the 
 BDMPS result (dot-dashed line). The coherent limits both for the Fokker-Planck
 approximation (dotted dark line) and the Debye potential (dotted light line)
 are also shown.}
\label{fig:figure_4}
\end{figure}
In the Gaussian/Fokker-Planck approximation for $\Sigma_2(\delta\v{k},\delta
l)$ we find in the coherence plateau where $\omega dI/d\omega$ is constant
analytical expressions for the asymptotic limits of very large and very small
mediums. An expression which interpolates between them can be found
\begin{equation}
i_0(l) \equiv \omega\frac{dI(l)}{d\omega}=\frac{2}{\pi}
 \alpha_sC_f\; \frac{1+\eta_c}{3 A + \eta_c}
       \log ( 1+A \eta_c ),
\label{coherence_plateau}
\end{equation}
where $A = e^{-(1+\gamma)}$, $\gamma$ is Euler's constant, and
$\eta_c=2\hat{q}l/m_g^2$ is a dimensionless number which is a measure of the
number of collisions.

In Figures \ref{fig:figure_1} and \ref{fig:figure_2} we show the results of 
our calculations for a medium density of $n_0 T_f$ = 8 fm$^{-3}$ and a gluon 
mass of $m_g$ = 0.45 GeV, for a medium length of $l =$ 1 fm and $l =$ 5 fm, respectively,
for both the Gaussian approximation and the Debye potential. We see that, for
the same parameters, the Debye potential produces more radiation. The
difference can be cast approximately into a redefinition of $\hat{q}$ but at
the cost of making it medium size, and Debye mass, dependent.  Also shown is our
estimation of the intensity using the approximated expression
(\ref{simplificada}) for the Gaussian case. As it can be seen, the
approximation is rather good, specially at large frequencies.  We show also
the result obtained by neglecting altogether the phases in the calculation
(labeled as coherent limit in the figure). In Figures \ref{fig:figure_3} and
\ref{fig:figure_4} we show the same results for a medium density of $n_0 T_f$
= 1 fm$^{-3}$ and a gluon mass of $m_g$ = 0.15 GeV and a medium length of $l
=$ 1 fm and $l$ = 5 fm, respectively.

\begin{figure}[h]
\includegraphics[scale=0.65]{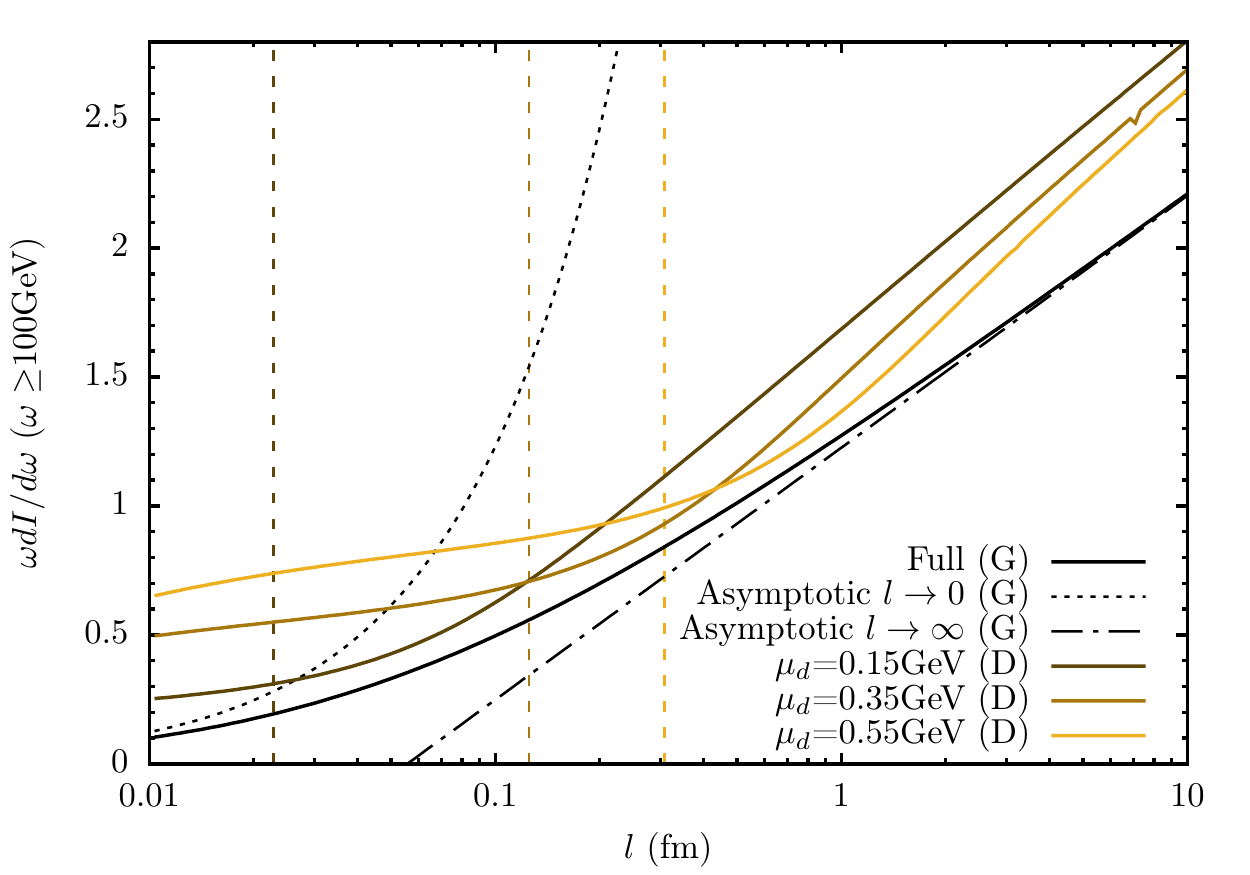}
\caption{Asymptotic emission intensity $i_0(l)$ as a function of the medium
  length for density $n_0 T_f = 8$ fm$^{-3}$ for the Gaussian approximation
  (continuous black line) and for the Debye potential for different Debye
  masses, as marked, $m_g = 0.45$ GeV. Also shown the large and small lengths
  approximations. Vertical dashed lines mark the
  transition between large and small media for each Debye mass, $\eta_c =2
  \hat{q}l/m_g^2 = 1$. }
\label{fig:figure_5}
\end{figure}

In Figure \ref{fig:figure_5} we show the asymptotic emission intensity
$i_0(l)$ as a function of the medium length for the Gaussian approximation and
for the Debye potential for different Debye masses (keeping the effective
$\hat{q}$ constant). As can be seen in the figure the ratio between the Debye
and the Gaussian intensities is not constant so that one can not redefine a
Gaussian $\hat{q}$ independently of the medium properties.
\begin{figure}[h]
\includegraphics[scale=0.65]{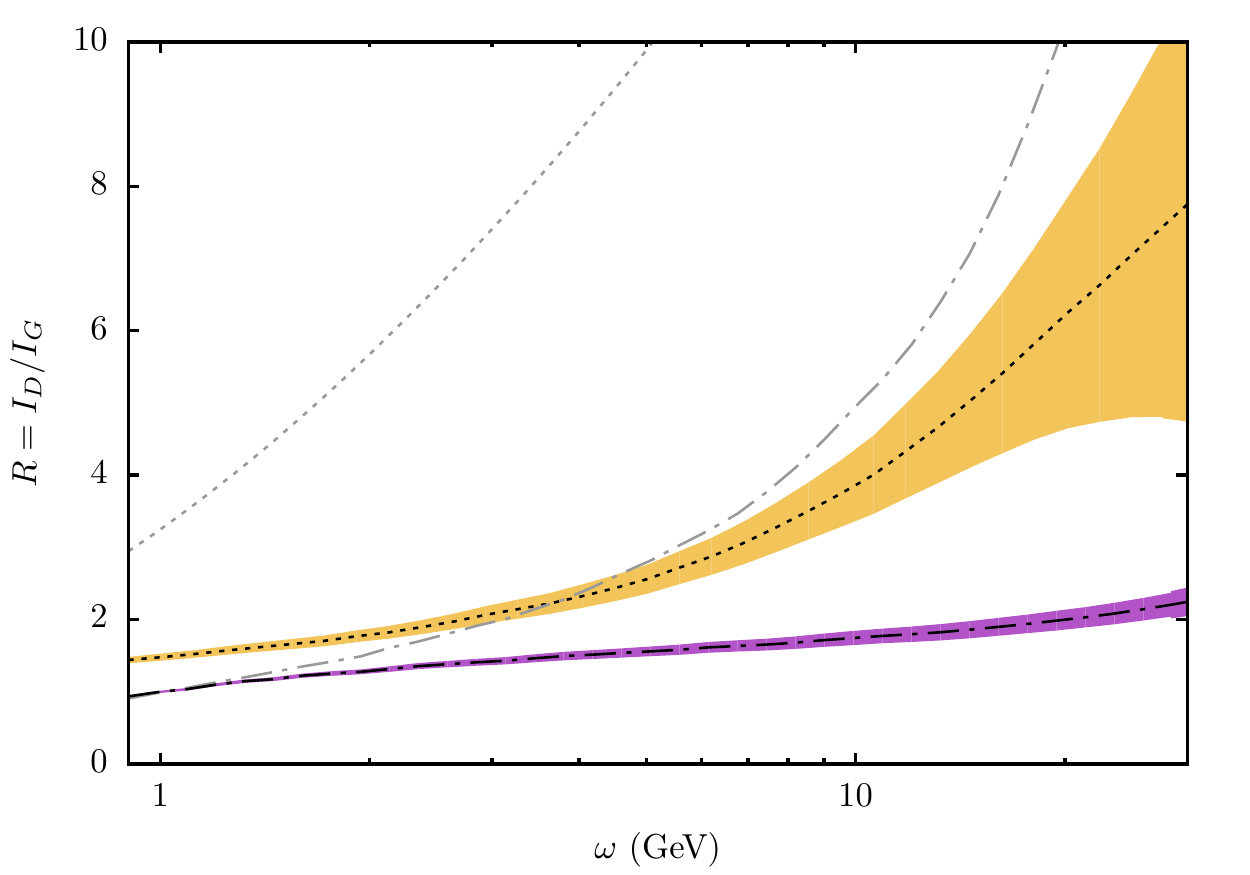}
\caption{Ratio of the bremsstrahlung intensity for the Debye potential
  compared to the Gaussian approximation for two gluon masses. We show the
  results for $n_0 T_f = 8 $ fm$^{-3}$, $\mu_d=0.45$ GeV and $l= 5$ fm as the
  dark dot-dashed line,  and $l= 1$ fm as the light dot-dashed line, 
  for $n_0 T_f = 1 $ fm$^{-3}$, $\mu_d=0.15$ GeV and $l= 5$ fm as the dark
  dotted line and $l=1$ fm as the light dotted line. Shadow bands show the
  statistical uncertainty. }
\label{fig:figure_6}
\end{figure}

\section{Radiation intensity after a first collision}
\label{interior_radiation}
We now consider a more realistic case where the parton suffers a first hard
collision.
We assume here that this scenario can be approximated within our formalism
taking into account gluons emitted after the first momentum change. This
corresponds, in the high energy limit, to a restriction of the $z$-integration
in (\ref{emission_amplitude_general}) to the interval $[0,+\infty)$. The
  square of this truncated amplitude can then be splitted into three terms
\begin{align}
\left|\int_0^{\infty}\right|^2= \int^l_0\int^l_0+
\int_l^{\infty}\int_l^{\infty}+2\Re\int^l_0\int_l^{\infty},
\end{align}
namely the intensity of a ``medium'' part, the intensity of the last leg, and
their interference. Since the last leg intensity appears now impaired in the
absence of a first leg amplitude, its integration in the gluon's transverse
momentum leads to a logarithmic divergence $\sim\log(\omega)$.
This divergence is removed, to make connection with the subset of terms
included in the BDMPS \cite{rbaier1995} calculations, by defining
\begin{align}
\left|M\right|^2 = \left|\int_0^{\infty}\right|^2-\left| \int_l^{\infty}
\right|^2= \int^l_0\int^l_0+2\Re\int^l_0\int_l^{\infty}
\label{squared_emission_amplitude_innerterms},
\end{align}  
which is interpretable as a probability only when $l\to\infty$,
corresponding in this limit to the semi-infinite medium evaluation of
(\ref{squared_emission_amplitude_allterms}).

In Figures \ref{fig:figure_1}, \ref{fig:figure_2}, \ref{fig:figure_3} and
\ref{fig:figure_4} the evaluation of
(\ref{squared_emission_amplitude_innerterms}) is shown together with the BDMPS
result \cite{dokshitzer2001,rbaier1997}.  Neither
(\ref{squared_emission_amplitude_innerterms}) nor the BDMPS calculation
consider the emission in the coherent limit, therefore the intensity goes to
zero for large frequencies. 
Whereas this prescription is correct for the infinite length limit, it is
assumed that for finite media the neglected terms are reabsorbed into the
structure and fragmentation functions. 
As can be seen in the figures, our results for the
Gaussian case agree with the BDMPS results at high energy, as expected. At low
energy, the kinematical restriction in the $k_t$ integration and the effect of
the mass of the gluon make the intensity to decrease.  As before, the
calculation done with the Debye potential gives a larger emission intensity
than the Gaussian case, for the same parameters. This can be seen in figure
\ref{fig:figure_6}, were we shown the ratio of the intensity for the Debye
calculation to the Gaussian calculation as a function of the energy of the
gluon emitted. This ratio is not a constant and depends on the mass of the
gluon going from $\sim 1$ ($\sim 1.5$) at low energies for $m_g=0.45$ GeV
($m_g=0.15$ GeV) to $\sim 2.2$ ($\sim 8$) at larger energies. Therefore a
change on the value of the Gaussian $\hat{q}$ can fit the Debye results but in
a qualitative way only. For instance a change of $\hat{q} \rightarrow 3.5
\hat{q}$ fits the result for the case $n_0 T_f = 8 $ fm$^{-3}$, $\mu_d=0.15$
GeV and large lengths ($l\gtrsim 3$ fm) with a maximum $20 \%$ error. For
Debye masses $\mu_d = 0.45$ GeV the scale factor is 2.8 with a larger error of
$\sim 40 \%$. In general, as expected, for small lengths the Debye result can
not be fit with a redefinition of $\hat{q}$.

\section{Conclusions}

We have developed a Monte Carlo method which is able to calculate the gluon
bremsstrahlung for realistic Debye screened interactions. A semi-analytical
approximation has been also estimated which helps to qualitatively understand
the LPM effect in QCD. The Fokker-Planck approximation is shown to
underestimate the emission intensity, and the difference can not be cast into
a redefinition of $\hat q$, independently of gluon's energy or medium size. If
we consider the total intensity (\ref{squared_emission_amplitude_allterms})
the finite size of the medium translates into a length-dependent coherent term
which would dominate the energy loss as $\Delta \propto E_q$ for large quark
energies. This was already found in \cite{vitev2007} in the case of cold
nuclear matter and applied to the azimuthal asymmetries in pA collisions in
\cite{gyulassy2014}. We found an expression for the intensity in the coherent
regime which confirms the same behavior in the small opacity limit, linear in
length and density.  In contrast, if we consider only the gluons emerging from
the inner legs (\ref{squared_emission_amplitude_innerterms}) energy loss is
dominated by the emission of small energy gluons and we recover the BDMPS
result except for kinematical restrictions and length effect.  In this
scenario one should expect a qualitatively different behavior since the
coherent plateau, related to the initial and final state radiation, is
missing.
A study of the full amplitude including the
interferences between the initial and final emissions was made in
\cite{Armesto}.  

In both scenarios we found that the emission is suppressed at
small gluon energies and is largest at around $\omega_c$ ($\sim$ 5-10 GeV) as
a result of medium incoherence.  The shape and depth of this intensity gives
information on the medium properties. A large width of the intensity is a sign
of a large length of the medium, whereas a narrow rise indicates a high
density medium or large gluon masses.

\acknowledgments

We thank N. Armesto, C.A. Salgado, and J. Sanchez-Guillen for helpful comments
and discussions. This work has been done in part under the grant Maria de
Maeztu Unit of Excellence (Spain).

\end{document}